\documentclass[12pt]{article}
\usepackage{ol2}
\usepackage{graphics}

\begin{document}

\title{Enhancement of semiconducting single-wall carbon nanotubes photoluminescence}

\author{Etienne Gaufr\`es$^1$, Nicolas Izard$^{2+}$, Laurent Vivien$^{1*}$, Sa\"id Kazaoui$^2$, Delphine Marris-Morini$^1$ and Eric Cassan$^1$}
\affiliation{1: Institut d'Electronique Fondamentale, CNRS-UMR 8622, Universit\'e Paris-Sud 11, 91405 Orsay, France}
\affiliation{2: National Institute of Advanced Industrial Science and Technology (AIST), 1-1-1 Higashi, Tsukuba, Ibaraki 305-8565, Japan}
+: Currently at Institut d'Electronique Fondamentale
\email{laurent.vivien@u-psud.fr}

\date{12/15/2009}

\begin{abstract} Photoluminescence properties of semiconducting single wall
	carbon nanotubes (s-SWNT) thin films with different metallic single wall
	carbon nanotubes (m-SWNT) concentrations are reported. s-SWNT purified
	samples are obtained by polymer assisted selective extraction. We show
	that a few m-SWNT in the sample generates a drastic quenching of the
	emission. Therefore, highly purified s-SWNT films are a strongly
	luminescent material and a good candidate for future applications in
	photonics, such as near infrared emitters, modulators and detectors.
\end{abstract} 


\maketitle

Single wall carbon nanotube (SWNT) is a unique monodimensional material,
exhibiting unusual electronic and optical properties, making it an ideal
candidate for future opto-electronic\cite{Avouris} and all-optical
devices\cite{Maruyama}. The nanotube structure, defined by the so-called (n,m)
chiral index, fixes all the nanotubes properties, such as metallic or
semiconducting behavior. Semiconducting nanotubes (s-SWNT) are of particular
interest in photonics for their direct band-gap, allowing near-IR luminescence
(PL) from electron-hole recombination. In the last few years, numerous
impressive results have been obtained on the photoluminescence properties of
SWNT, where luminescence arise from well-isolated s-SWNT, either in solution by
wrapping with surfactant\cite{Oconnel, Bachilo, Arnold-nanolett}, or dispersion
in gelatin matrix\cite{Wang, Berger}. Recently, photoluminescent composite gels
have also been demonstrated\cite{Zamora}. Indeed, s-SWNT are extremely sensitive
to their environment. An incomplete isolation, resulting in small bundles or
metallic nanotubes (m-SWNT) in contact with s-SWNT, will lead to the creation of
alternative non-radiative de-excitation paths and ultimately an effective
quenching of the photoluminescence. It has been estimated that only around 10 \%
of the s-SWNT embedded in gelatin matrix effectively participate in the
fluorescence, the remainder being quenched by left-over m-SWNT, bundles or
impurities\cite{Berger}. The direct influence of m-SWNT removal on the nanotubes
optical properties was never really studied. Several approaches have been
explored in the past few years for optimizing the diameter and chirality
selection of SWNTs, such as chemical functionalization, DNA, polymer wrapping
and density gradient centrifugation techniques\cite{Menard, Zhang,
Arnold-Nature, Nish, Chen, Hwang}. The possibility to achieve chiral specific
SWNT extraction is indeed very promising for future photonic applications.
Recently, we demonstrated that the use of poly-9,9-di-n-octyl-fluorenyl-2,7-diyl
(PFO) in toluene as an extracting agent followed by ultracentrifugation allows
to obtain quasi metallic-free SWNT\cite{Izard}. Using sample with a constant
concentration in s-SWNT but differents concentrations in m-SWNT, we show in this
paper that the presence of even a few metallic nanotubes quenche the strong
intrinsic photoluminescence properties of s-SWNT.

\begin{figure}
	\includegraphics[width=15cm]{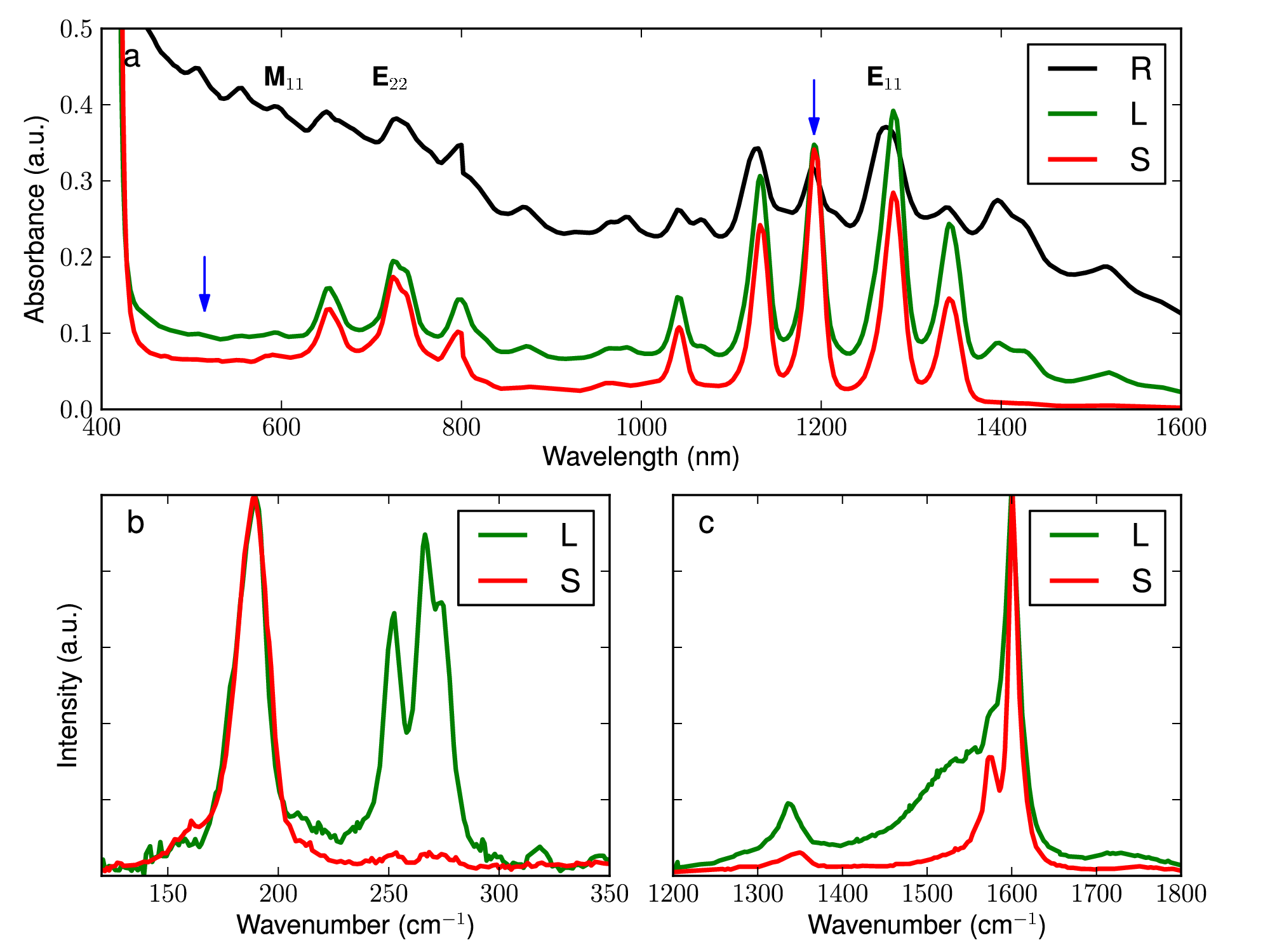}
	\caption{(color online) (a) The optical absorption
	spectra in toluene, (b) and (c) the RBM and TM Raman spectra at 2.41~eV
	of R, L and S samples. Arrows indicates respectively position of the
	Raman laser probe (left) and  where the concentration adjustment was
	performed (right).\label{fig1}}
\end{figure}

Samples were prepared as previously described\cite{Izard}. First, SWNT powder
(as-prepared HiPco, Carbon Nanotechnologies Inc.),
poly-9,9-di-n-octyl-fluorenyl-2,7-diyl (PFO) (Sigma-Aldrich) and toluene were
mixed in the ratio of SWNT (5mg):PFO (20mg):toluene (30ml) and homogenized by
sonication (for 1 h using a water-bath sonicator and 15 min using a tip
sonicator). Then this mixture was centrifugated for 5-60 min using either a
desktop centrifuge (angle rotor type, 10.000g) or an ultracentrifuge (swing
rotor type, 150.000g), after which the upper 80 \% of the supernatant solution
was collected. We shall focus on two types of SWNT solution: solution
centrifugated at 10.000g for 15 min (labeled L) and solution centrifugated at
150.000 g for 2 hours (labeled S). We also consider the non centrifugated
solution (labeled R) for comparison purpose. At this stage, the concentration of
solutions R and L were adjusted by dilution to match the most intense absorption
peak around 1200~nm (1.03~eV) of sample S to assure the same amount of s-SWNTs
in all off the solutions (cf. Figure \ref{fig1}a). Solutions were then spin
coated several times onto glass substrate at 500~RPM for 60~s to achieve a layer
thickness around 200~nm\cite{thickness}. Lastly, samples were annealed at
150~$^{o}$C for 15~min.

Absorption spectra of R, L and S samples in toluene are presented in Figure
\ref{fig1}a. Sharp peaks in the range of 1050-1400 nm (1.18-0.89~eV, labeled
E$_{11}$) and 600-900 nm (2.07-1.38~eV, E$_{22}$) are the optical absorption
bands corresponding respectively to the first and second transitions between
Van Hove singularities in the s-SWNT density of states. Peaks in the range of
500-600 nm (2.48-2.07~eV, M$_{11}$) correspond to the absorption bands of
m-SWNTs\cite{Chiang}. Peak intensities gradually diminished and eventually
vanished when all the m-SWNTs were removed by ultracentrifugation in sample S.
The absorption background also drastically reduced after the first low gravity
(G) centrifugation. Indeed, the centrifugation step permits the effective
removal of amorphous carbon and catalyst impurities initially present in the
HiPCO SWNT powder. Raman spectra for samples L and S were recorded at 514.5~nm
(2.41~eV) and the radial breathing mode (RBM) and tangential mode (TM) regions
are reported in Figure \ref{fig1} b and c, respectively. Raman spectroscopy is
a resonant process for SWNT, and the RBM peak at 190~cm$^{-1}$ is specific to
s-SWNT while the RBM peak at 270~cm$^{-1}$ is specific to
m-SWNT\cite{Sauvajol}. We observed that m-SWNT peaks present in sample L
completely vanished in sample S. This is coherent with the TM regions, where
the broad asymmetric line shape between 1400 and 1600 cm$^{-1}$ of sample L,
characteristic of a metallic behavior, completely disappears in sample S.
Further electric measurements presented in reference \cite{Izard} confirm the
absence of m-SWNT in sample S. According to absorption and Raman measurements,
the first low G centrifugation step (sample L) effectively remove the
initially present amorphous carbon and catalyst impurities, but also remove
some m-SWNT.  The high G centrigugation step (sample S) remove all the
remaining m-SWNT. Thus, the composition of samples L and S differ only by the
presence of m-SWNT in sample L.

\begin{figure}
	\includegraphics[width=15cm]{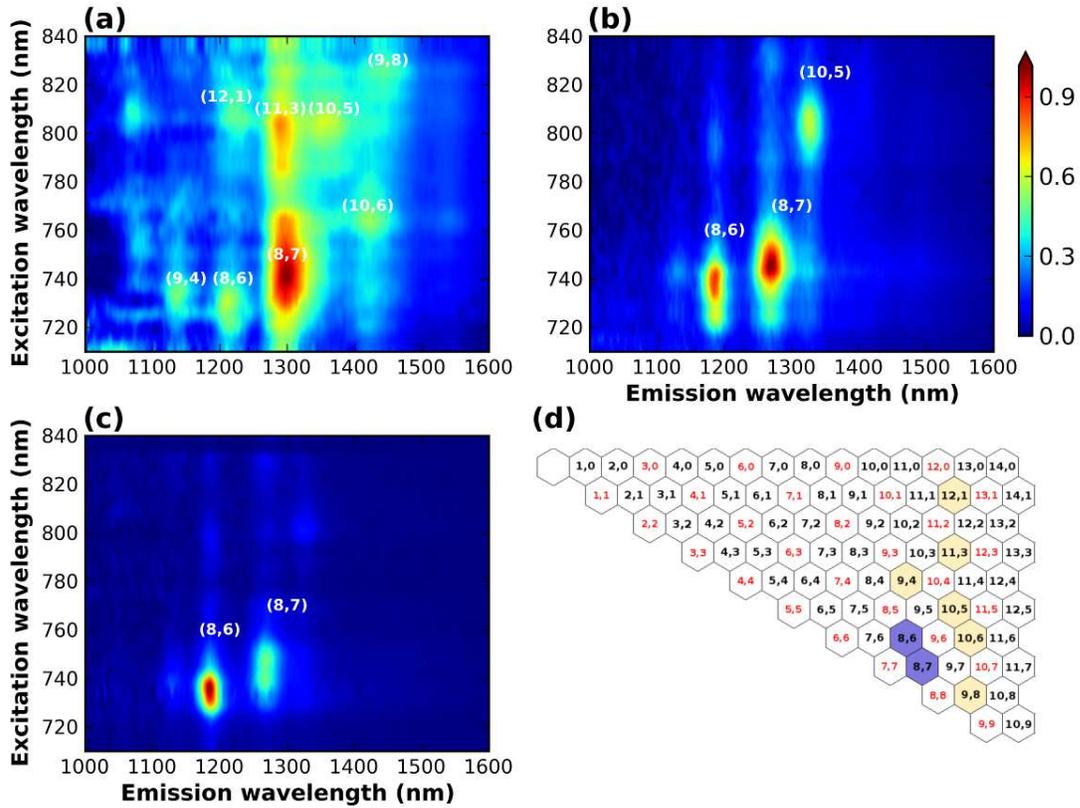}
	\caption{(color online) Photoluminescence map
	(emission wavelength vs excitation wavelength) of PFO- embedded SWNT
	thin film spin-coated onto glass substrate. The layer thickness is
	around 200 nm. (a) Sample R, (b) sample L and (c) sample S. (d) Chiral
	map showing the wrapping preference of PFO (in blue / dark gray) as compared to raw
	HiPCO SWNT sample (in yellow / light gray).
	\label{fig2}}
\end{figure}

The PL excitation-emission maps of each samples were obtained using a Titane
Sapphire (Ti:Sa) laser pumped by a cw Ar laser. Ti:Sa laser delivers
continuous light at a wavelength range from 700~nm to 840~nm (1.77 to
1.48~eV). Photoluminescence signal is then recorded at room temperature using
a monochromator (JobinYvon 550) coupled with a cooled InGaAs detector with a
3~nm step. Emission intensity as a function of both excitation (from 700 to
840 nm / 1.77 to 1.48~eV) and emission wavelength (from 1000 to 1600 nm / 1.24
to 0.78~eV) was recorded. The experimental values were corrected by taking
into account the system response. Results are presented in Figure \ref{fig2}a,
b and c, each map being normalized by the maximal intensity. Figure
\ref{fig2}d shows the chirality map for the photoluminescent nanotubes from
raw nanotubes in sample R and PFO purified s-SWNT in sample S. We discovered,
as previously observed by Chen et al.\cite{Chen}, a chiral selectivity of PFO
polymer. Indeed, chirality in the starting material were evenly distributed
inside the measuring range, with chiral angle ranging from 3.96$^\circ$ in
(12,1) tube to 27.8$^\circ$ in (8,7) tubes. In contrast, PFO-extracted
nanotubes consist of only two tubes in the measuring range, 73 \% of (8,6) and
27 \% of (8,7) nanotubes\cite{calcul}. Those two nanotubes possess near
armchair structure with high chiral angles above 25$^\circ$. The exact
mechanism of PFO's high chiral selectivity is still unknown, but the aromatic
structure of the polymer may play a major role in the wrapping interaction
with nanotubes\cite{Hwang}.

\begin{figure}
	\includegraphics[width=15cm]{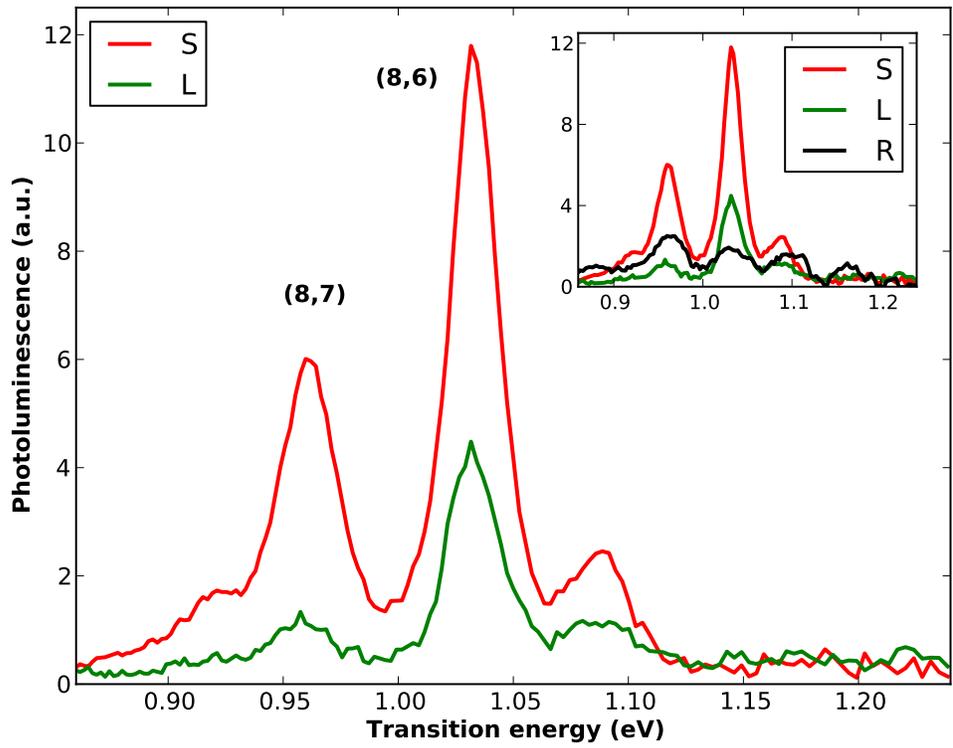}
	\caption{(color online) Photoluminescence intensities of samples L and S which differ in m-SWNT
	concentration.  Intensity was normalized taking into account the net
	intensity of light absorbed in s-SWNT at the pumping enegy (1.7 eV).
	Inset: Photoluminescence intensities of L and S samples compared to R
	sample, without normalization.
	\label{fig3}}
\end{figure}

In order to determine the influence of the purification process, SWNT
photoluminescence intensity for an excitation energy of 1.7 eV is displayed in
Figure \ref{fig3}. Particular care was given to obtain the three samples with
roughly the same concentration of (8,6) s-SWNT as shown on Figure \ref{fig1}a,
which permit a comparison of the photoluminescence intensity level.
Furthermore, photoluminescence intensity of S and L samples were corrected by
the absorption peak intensity at 1.7 eV after background substraction. It was
not possible to properly remove all background contribution in R sample;
uncorrected photoluminescence intensities of all three samples are presented
in the inset of Figure \ref{fig3}. The most striking feature is the impressive
increase of (8,6) nanotube photoluminescence intensity with the degree of
purification. Indeed, while sample R presents a rather low signal, samples L
and S exhibit PL signal enhancement. An increase by a factor of 3 is observed
in the PL intensity of (8,6) nanotube for sample S as compared to L.  PL
intensities of L and S samples increase by a factor of 2 and 6 respectively
compared to sample R (unpurified). An analog increase was also observed for
the intensity of the (8,7) nanotube, but as the relative concentration in
(8,7) nanotube change between samples, the comparison is not as
straightforward. It is remarkable that even if the concentration of (8,7) tube
is higher in the starting material, the photoluminescence intensity is still 3
times higher in the PFO selectively extracted material. As the number of
quenching centers constituted by m-SWNT decreases, most of the exciton
recombination occurring in the (8,6) nanotubes likely contribute to the
luminescence phenomenon, leading to an intensity increase of S sample by
factor of 3 and 6 as compared to L and R samples, respectively.

In conclusion, we have studied the photoluminescence of pure semiconducting
SWNT thin films, extracted by the PFO centrifugation technique. We found that
the removal of m-SWNTs leads to an enhancement of the photoluminescence
properties. Indeed, after removal of m-SWNT, the same quantity of (8,6) s-SWNT
displays a 6-fold increase in photoluminescence intensity. The use of PFO as a
matrix leads to the formation of well isolated nanotube thin films. These
results are of major importance for the future use of s-SWNT thin film matrix
in photonic applications.

N. Izard thanks the Japan Society for the Promotion of Science for financial support.

\pagebreak

\pagebreak

\pagebreak

\pagebreak


\begin{thebibliography}{00}
\bibitem{thickness}
Layer thickness was estimated from ellipsometry spectroscopy and direct profilometer measurement.
\end{thebibliography}
\end{document}